\documentclass[10pt]{article}
\usepackage{epsfig}
\usepackage{amsmath}
\newcommand{\ga}{\gamma}
\newcommand{\varep}{\varepsilon}
\newcommand{\la}{\lambda}
\newcommand{\varph}{\varphi}
\newcommand{\ps}{\psi}
\newcommand{\rh}{\rho}
\newcommand{\si}{\sigma}
\newcommand{\vac}[1]{\mbox{\boldmath$\langle$}{#1}\mbox{\boldmath$\rangle$}}
\newcommand{\klgl}{\:\hbox to -0.2pt{\lower2.5pt\hbox{$\sim$}\hss}
 {\raise3pt\hbox{$<$}}\:}
\newcommand{\fm}{{\rm \;fm}}
\newcommand{\GeV}{{\rm \;GeV}}
\newcommand{\nn}{\nonumber}
\newcommand{\figepsSC}[2]{\epsfxsize=#1mm\epsfbox[72 240 540 540]{./#2.eps}
  \setlength{\unitlength}{0.464mm}}
\newcommand{\figpsSC}[2]{\epsfxsize=#1mm\epsfbox[100 300 530 570]{./#2.ps}
  \setlength{\unitlength}{0.464mm}}
\begin{document}
\bibliographystyle{personale}
\sloppy

\vspace*{-2.5cm}
\begin{center} { HD-THEP-99-1. -- {\it To be published in} {\rm Fizika} {\bf B} } \end{center}
\vspace*{.5cm}

\begin{center}
{\large VECTOR MESONS $\rh$, $\rh'$ AND $\rh''$ \\
  DIFFRACTIVELY PHOTO- AND LEPTOPRODUCED } \\ \medskip
\makeatletter
{G.~KULZINGER
  \footnote{
  Supported by the Deutsche Forschungsgemeinschaft under grant no. GRK 216/1-96
  }$^,$\footnote{E-mail: G.Kulzinger@thphys.uni-heidelberg.de} }
\makeatother \\ \medskip
{\it Institut f\"ur Theoretische Physik der Universit\"at Heidelberg, \\
Philosophenweg 16, 69120 Heidelberg, Germany } \\
\end{center} \medskip
\noindent
In the framework of non-perturbative QCD we calculate high-energy diffractive production of vector mesons $\rh$, $\rh'$ and $\rh''$ by real and virtual photons on a nucleon. The initial photon dissociates into a $q\bar{q}$-dipole and transforms into a vector meson by scattering off the nucleon which, for simplicity, is represented as quark-diquark. The relevant dipole-dipole scattering amplitude is provided by the non-perturbative model of the stochastic QCD vacuum. The wave functions result from considerations in the frame of light-front dynamics; the physical $\rh'$- and $\rh''$-mesons are assumed to be mixed states of an active $2S$-excitation and some residual rest ($2D$- and/or hybrid state). We obtain good agreement with the experimental data and get an understanding of the markedly different $\pi^+\pi^-$-mass spectra for photoproduction and $e^+e^-$-annihilation. \medskip

\noindent
Keywords: non-perturbative QCD, diffraction, photoproduction, photon wave function, $\rh$-meson, excited vector mesons, hybrid
\bigskip

\begin{center} {\large\it 1. Introduction} \end{center}
Diffractive scattering processes are characterized by small momentum transfer, $-t\klgl 1$~GeV$^2$, and thus governed by non-perturbative QCD. To get more insight in the physics at work we investigate exclusive vector meson production by real and virtual photons. In this note we summarize recent results from Ref.~\cite{KDP} on $\rh$-, $\rh'$- and $\rh''$-production, see also Ref.~\cite{K}. In Refs~\cite{KDP,DGKP} we have developed a framework which we here can only flash.

We consider high-energy diffractive collision of a photon, which dissociates into a $q\bar{q}$-dipole and transforms into a vector meson, with a proton in the quark-diquark picture, which remains intact. The scattering $T$-amplitude can be written as an integral of the dipole-dipole amplitude and the corresponding wave functions. Integrating out the proton side, we have
\begin{equation} \label{T_amplitude}
T^\la_V(s,t) = is \int \frac{dzd^2{\bf r}}{4\pi}\,
  \psi^\dagger_{V(\la)}\psi_{\ga(Q^2,\la)}(z,{\bf r})\;
  J_p(z,{\bf r},\Delta_T) \;,
\end{equation}
where $V(\la)$ stands for the final vector meson and $\ga(Q^2,\la)$ for the initial photon with definite helicities $\la$ (and virtuality $Q^2$); $z$ is the light-cone momentum fraction of the quark, $\bf r$ the transverse extension of the $q\bar{q}$-dipole. The function $J_p(z,{\bf r},\Delta_T)$ is the interaction amplitude for a dipole $\{z,{\bf r}\}$ scattering on a proton with fixed momentum transfer $t\!=\!-\Delta_T^2$; for $\Delta_T\!=\!0$ due to the optical theorem it is the corresponding total coss section (see below Eq.~(\ref{sigma_dipole})). It is calculated within non-perturbative QCD: In the high-energy limit Nachtmann~\cite{Na} derived a non-perturbative formula for dipole-dipole scattering whose basic entity is the vacuum expectation value of two lightlike Wilson loops. This gets evaluated~\cite{DFK} in the model of the stochastic QCD vacuum.

\begin{center} {\large\it 2. The model of the stochastic QCD vacuum} \end{center}
Coming from the functional integral approach the model of the stochastic QCD vacuum~\cite{DoSi} assumes that the non-perturbative part of the gauge field measure, i.e. long-range gluon fluctuations that are associated with a non-trivial vacuum structure of QCD, can be approximated by a stochastic process in the gluon field strengths with convergent cumulant expansion. Further assuming this process to be gaussian one arrives at a description through the second cumulant $\vac{g^2 F_{\mu\nu}^A(x;x_0) F_{\rh\si}^{A'} (x';x_0)}$ which has two Lorentz tensor structures multiplied by correlation functions $D$ and $D_1$, respectively. $D$ is non-zero only in the non-abelian theory or in the abelian theory with magnetic monopoles and yields linear confinement. Whereas the $D_1$-structure is not confining.

\begin{figure}
$$
\figpsSC{65}{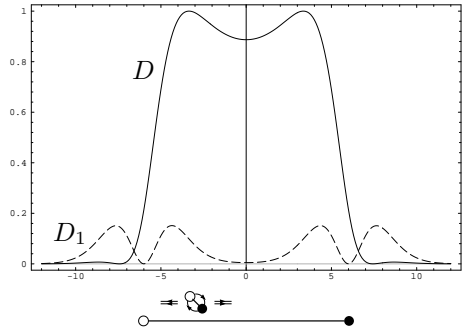}
\begin{picture}(0,0)
\put(-124,24){\makebox(0,0){$D_1$}}
\put(-103,71){\makebox(0,0){$D$}}
\end{picture}
$$
\vspace*{-3ex}
\caption[]{\small Interaction amplitude (arbitrary units) of two colour dipoles as function of their impact (units of correlation lengths $a$). One large $q\bar{q}$-dipole of extension $12a$ is fixed, the second small one of extension $1a$ is, averaged over all its orientations, shifted along on top of the first one. For the $D_1$-tensor structure of the correlator there are only contributions when the endpoints are close to each other, whereas for the $D$-structure large contributions show up also from between the endpoints. This is to be interpreted as interaction with the gluonic string between the quark and antiquark.} \label{Fig:dipdip}
\end{figure}

\begin{figure}[!t]
$$
\figpsSC{65}{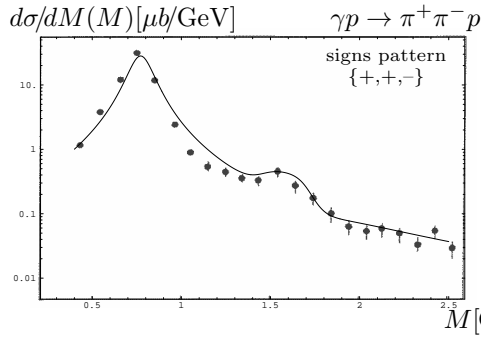}
\begin{picture}(0,0)
\put(-32,79){\makebox(0,0){\scriptsize signs pattern}}
\put(-32,73){\makebox(0,0){\scriptsize \{+,+,--\}}}
\put(-2,3){\makebox(0,0){$M[\!\!\GeV]$}}
\put(-140,90){\makebox(135,0){$d\si\!/\!dM(M)[\mu b\!/\!\!\GeV]$
    \hfill $\ga p\rightarrow\pi^+\pi^-p$}}
\end{picture}
$$
\vspace*{-7ex}
\caption[]{Mass spectrum of $\pi^+\pi^-$-photoproduction on the proton: The interference in the 1.6 GeV region is constructive in contrary to the case of $e^+e^-$-annihilation into $\pi^+\pi^-$. We display our calculation together with the experimental data~\cite{DoMi}.} \label{Fig:spectra}
\end{figure}

\begin{figure}[!b]
$$
\figepsSC{65}{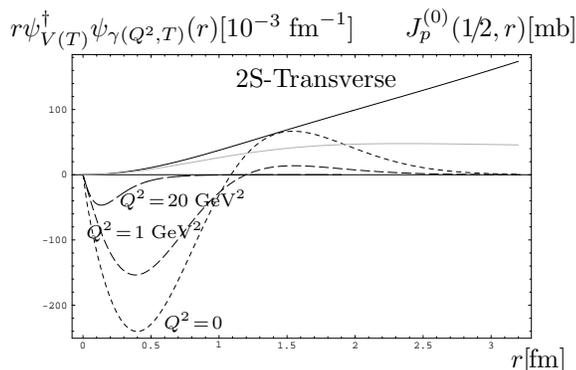}
\begin{picture}(0,0)
\put(2,1){\makebox(0,0){$r[\!\!\fm]$}}
\put(-152,91){\makebox(165,10){
    $r\ps_{V(T)}^\dagger\ps_{\ga(Q^2,T)}(r)[10^{-3}\fm^{-1}]$
    \hfill $J_p^{(0)}(1\!/\!2,r)[\rm mb]$}}
\put(-62,82){\makebox(0,0){2S-Transverse}}
\put(-97,12){\makebox(0,0){\scriptsize $Q^2\!=\!0$}}
\put(-111,38){\makebox(0,0){\scriptsize $Q^2\!=\!1\GeV^2$}}
\put(-100,47){\makebox(0,0){\scriptsize $Q^2\!=\!20\GeV^2$}}
\end{picture}
$$
\vspace*{-6ex}
\caption[]{\small Dipole-proton total cross section $J_p^{(0)}$ and the effective overlap $r\ps_{V(T)}^\dagger\ps_{\ga(Q^2,T)}$ as function of the transverse dipole size $r$. The black line is the function $J_p^{(0)}(1\!/\!2, r)$, i.e. the total cross section of a $q\bar{q}$-dipole $\{z\!=\!1\!/\!2,{\bf r}\}$, averaged over all orientations, scattering on a proton; the grey line shows the cross section for a completely abelian, non-confining theory. The $T$-amplitude is obtained by integration over the product of $J_p$ and the overlap function, which essentially is the effective overlap shown for $Q^2\!=\!0,\;1$ and 20~GeV$^2$ as short, medium and long dashed curves, respectively.} \label{Fig:overlap2S}
\end{figure}

The underlying mechanism of (interacting) gluonic strings also shows up in the scattering of two colour dipoles, cf.~Fig.~\ref{Fig:dipdip}, and essentially determines the $T$-amplitude if large dipole sizes are not suppressed by the wave functions. To confront with experiment this specific-large distance prediction we are intended to study the broad $\rh$-states and, especially, their production by broad small-$Q^2$ photons. Before we enter the discussion of our results, however, we have to specify these states and have to fix their wave functions as well as that of the photon.
%
%
\begin{center} {\large\it 3. Physical states $\rh$, $\rh'$ and $\rh''$} \end{center}
Analyzing the $\pi^+\pi^-$-invariant mass spectra for photoproduction and $e^+e^-$-anni\-hilation Donnachie and Mirzaie~\cite{DoMi} concluded evidence for two resonances in the 1.6~GeV region whose masses are compatible with the $1^{--}$ states $\rh(1450)$ and $\rh(1700)$. We make as simplest ansatz
\begin{eqnarray}
|\rh(770)\rangle\;\, &=& \;\;\; |1S\rangle \;, \\
|\rh(1450)\rangle    &=& \;\;\; \cos\theta \; |2S \rangle
                               + \sin\theta \; |rest \rangle \;, \nn \\
|\rh(1700)\rangle    &=&       -\sin\theta \; |2S \rangle
                               + \cos\theta \; |rest \rangle \;, \nn
\end{eqnarray}
where $|rest\rangle$ is considered to have $|2D\rangle$- and/or hybrid components whose couplings to the photon both are suppressed, see Refs.~\cite{BST} and~\cite{ClDo,ClPa}, respectively. With our convention of the wave functions the relative signs $\{+,-,+\}$ of the production amplitudes of the $\rh$-, $\rh'$- and $\rh''$-states in $e^+e^-$-annihilation determine the mixing angle to be in the first quadrant; from~Ref.~\cite{DoMi} then follows $\theta\!\cong\!41^\circ$. With this value and the branching ratios of the $\rh'$- and $\rh''$-mesons into $\pi^+\pi^-$ extracted in Ref.~\cite{DoMi} we calculate the photoproduction spectrum as shown in Fig.~\ref{Fig:spectra} with the observed signs pattern $\{+,+,-\}$; for details cf.~\cite{KDP}. We will understand below from Fig.~\ref{Fig:overlap2S} the signs change of the $2S$-production as due to the dominance of large dipole sizes in photoproduction in contrary to the coupling to the electromagnetic current $f_{2S}$ being determined by small dipole sizes.
%
%
\begin{center} {\large\it 4. Light-cone wave functions} \end{center}
In the high-energy limit the photon can be identified as its lowest Fock, i.e. $q\bar{q}$-state. The vector meson wave function distributes this $q\bar{q}$-dipole~$\{z,{\bf r}\}$,~accordingly.

{\bf Photon:} With mean of light-cone perturbation theory (LCPT) we get explicit expressions for both longitudinal and transverse photons. The photon transverse size which we will see to determine the $T$-amplitude is governed by the product $\varep r$,~$\varep \!=\! \sqrt{z\bar{z} Q^2 \!+\! m^2}$ and~$r\!=\!|{\bf r}|$. For high $Q^2$ longitudinal photons dominate by a power of $Q^2$; their $z$-endpoints being explicitly suppressed, LCPT is thus applicable. For moderate $Q^2$ also transverse photons contribute which have large extensions because endpoints are not suppressed. For $Q^2$ smaller than 1~GeV$^2$ LCPT definitively breaks down. However, it was shown~\cite{DGP} that a quark mass phenomenologically interpolating between a zero valence and a $220$~MeV constituent mass astonishingly well mimics chiral symmetry breaking and confinement. Our wave function is thus given by LCPT with such a quark mass $m(Q^2)$, for details cf. Refs~\cite{KDP,DGKP}.

{\bf Vector mesons:} The vector mesons wave functions of the $1S$- and $2S$-states are modelled according to the photon. We only replace the photon energy denominator $(\varep^2\!+\!{\bf k}^2)^{-1}$ by a function of $z$ and $|{\bf k}|$ for which ans\"atze according to Wirbel and Stech~\cite{WSB} are made; for the "radial" excitation we account by both a polynomial in $z\bar{z}$ and the $2S$-polynomial in ${\bf k}^2$ of the transverse harmonic oscillator. The parameters are fixed by the demands that the $1S$-state reproduces $M_\rh$ and $f_\rh$ and the $2S$-state is both normalized and orthogonal on the $1S$-state. For details cf. Ref.~\cite{KDP}.

\begin{center} {\large\it 5. Results} \end{center}
Before presenting some of our results~\cite{KDP} we stress that all calculated quantities are absolute predictions. Due to the eikonal approximation applied, the cross sections are constant with total energy $s$ and refer to $\sqrt{s}\!=\!20$~GeV where the proton radius is fixed. (The two parameters of the model of the stochastic QCD vacuum, the gluon condensate $\vac{g^2FF}$ and the correlation length $a$, are determined by matching low-energy and lattice results, cf. Ref.~\cite{DFK}.)

In Fig.~\ref{Fig:overlap2S} we display~-- for the transverse $2S$-state,~$\la\!=\!T$~-- both the functions
\begin{eqnarray}
J_p^{(0)}(z,r)
  &:=& \int_0^{2\pi} \frac{d\varph_{\bf r}}{2\pi}\,J_p(z,{\bf r},\Delta_T=0)
  \\ \label{sigma_dipole}
r\ps_{V(\la)}^\dagger\ps_{\ga(Q^2,\la)}(r)
&:=& \int \frac{dz}{4\pi}\; \int_0^{2\pi} \frac{d\varph_{\bf r}}{2\pi}\;
       |{\bf r}|\,\ps_{V(\la)}^\dagger\ps_{\ga(Q^2,\la)}(z,{\bf r})
\end{eqnarray}
which together, see Eq.~(\ref{T_amplitude}), essentially determine the leptoproduction amplitude. 
It is strikingly shown how for decreasing virtuality~$Q^2$ the outer positive region of the wave functions effective overlap $r\ps_{V(\la)}^\dagger\ps_{\ga(Q^2,\la)}$ wins over the inner negative part due to the strong rise with $r$ of the dipole-proton interaction amplitude $J_p^{(0)}$ which itself is a consequence of the string interaction mechanism discussed above. {\it In praxi\/} dipole sizes up to $2.5$~fm contribute significantly to the cross section.
%
%

\begin{figure}[!t]
$$
\figepsSC{65}{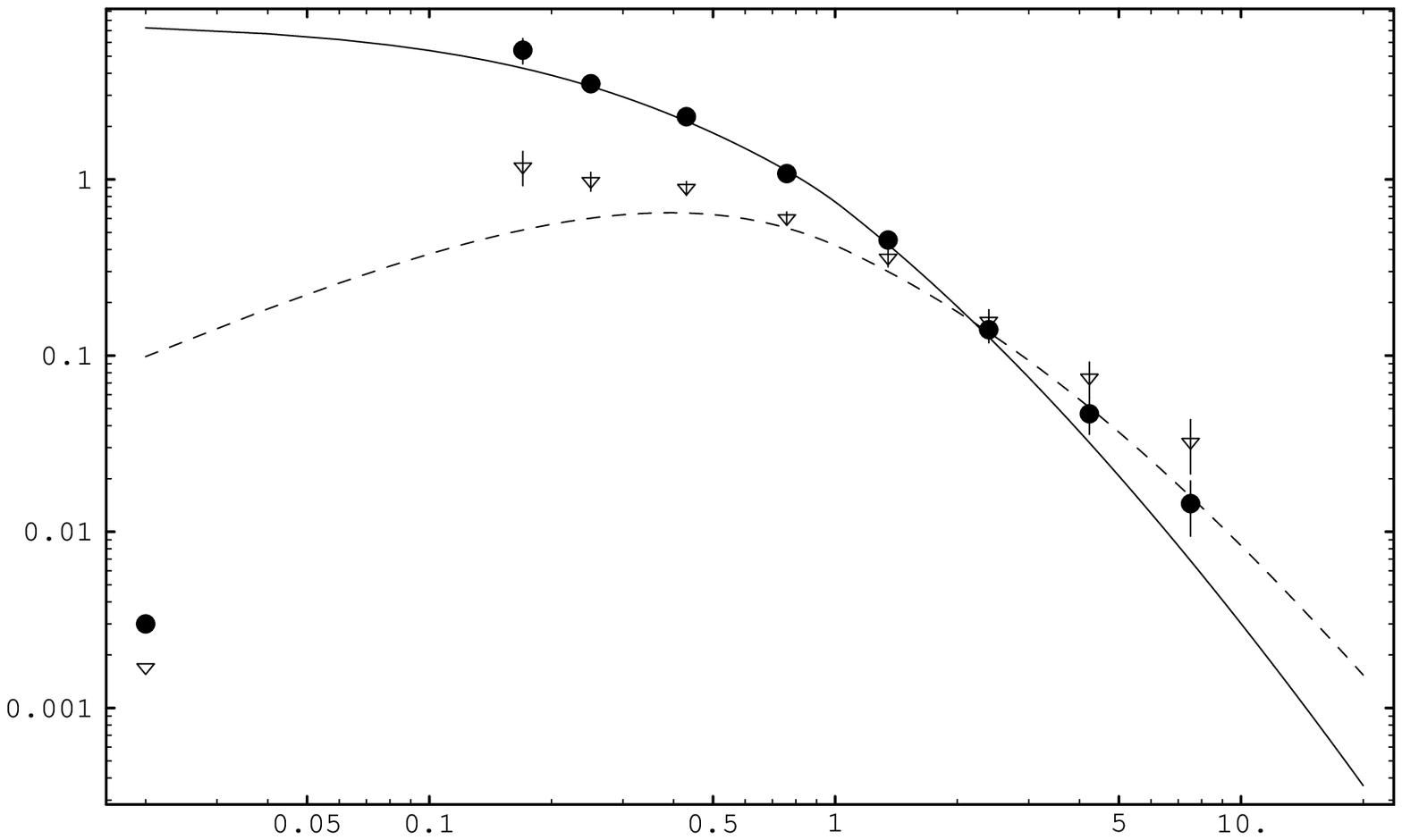}
\begin{picture}(0,0)
\put(-6,0){\makebox(0,0){$Q^2[\!\!\GeV^2]$}}
\put(-127,94){\makebox(0,0){$\si(Q^2)[\rm \mu b]$}}
\put(-109,80){\makebox(0,0){\scriptsize $\rh$-Transverse}}
\put(-106,50){\makebox(0,0){\scriptsize $\rh$-Longitudinal}}
\put(-96,27){\makebox(0,0){\scriptsize 85\% E665-Transverse}}
\put(-93,22){\makebox(0,0){\scriptsize 85\% E665-Longitudinal}}
\end{picture}
$$
\vspace*{-2ex}
$$
\figepsSC{65}{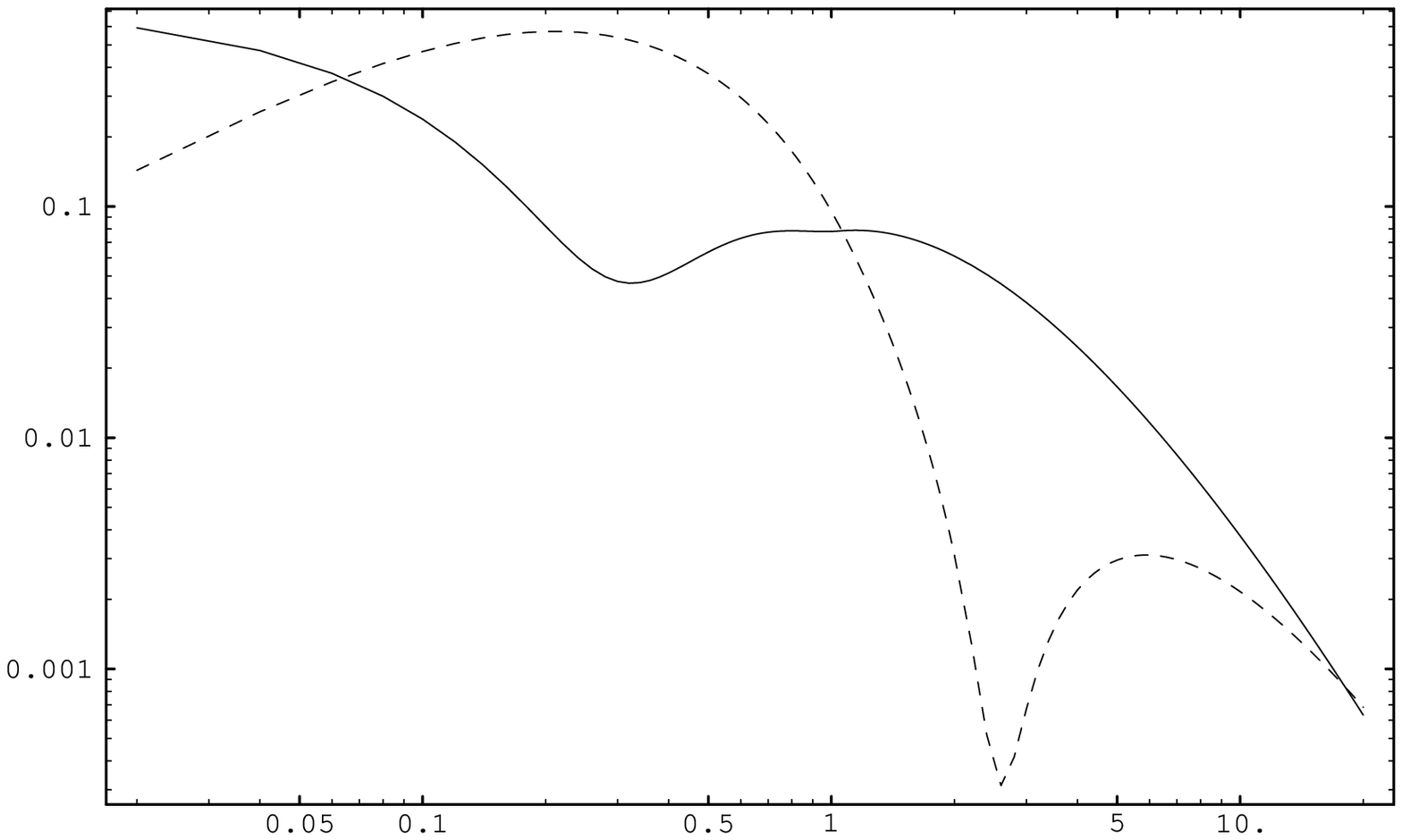}
\begin{picture}(0,0)
\put(-6,0){\makebox(0,0){$Q^2[\!\!\GeV^2]$}}
\put(-127,94){\makebox(0,0){$\si(Q^2)[\rm \mu b]$}}
\put(-40,80){\makebox(0,0){\scriptsize 2S-Longitudinal}}
\put(-21,65){\makebox(0,0){\scriptsize 2S-Transverse}}
\end{picture}
$$
\vspace*{-6ex}
\caption[]{\small Integrated cross sections of the $\rh$-meson and the $2S$-state as a function of $Q^2$. E665~\cite{E665_1} provides data for the $\rh$; the pomeron contribution which corresponds to our calculation we roughly estimate as 85\% of the measured cross section, cf. Ref.~\cite{DoLa}. } \label{Fig:sigma}
\end{figure}

Our results for integrated elastic cross sections as functions of $Q^2$ are given in Fig.~\ref{Fig:sigma}. For the $\rh$-meson our prediction is about $20-30\%$ below the E665-data~\cite{E665_1}. However, we agree with the NMC-experiment~\cite{NMC} which measures some definite superposition of longitudinal and transverse polarization, see Table~3 in Ref.~\cite{KDP}. For the $2S$-state, due to the nodes of the wave function, we predict a marked structure; the explicit shape, however, strongly depends on the parametrization of the wave functions.
%
%
%

%
%
\begin{figure}[!b]
$$
\figepsSC{65}{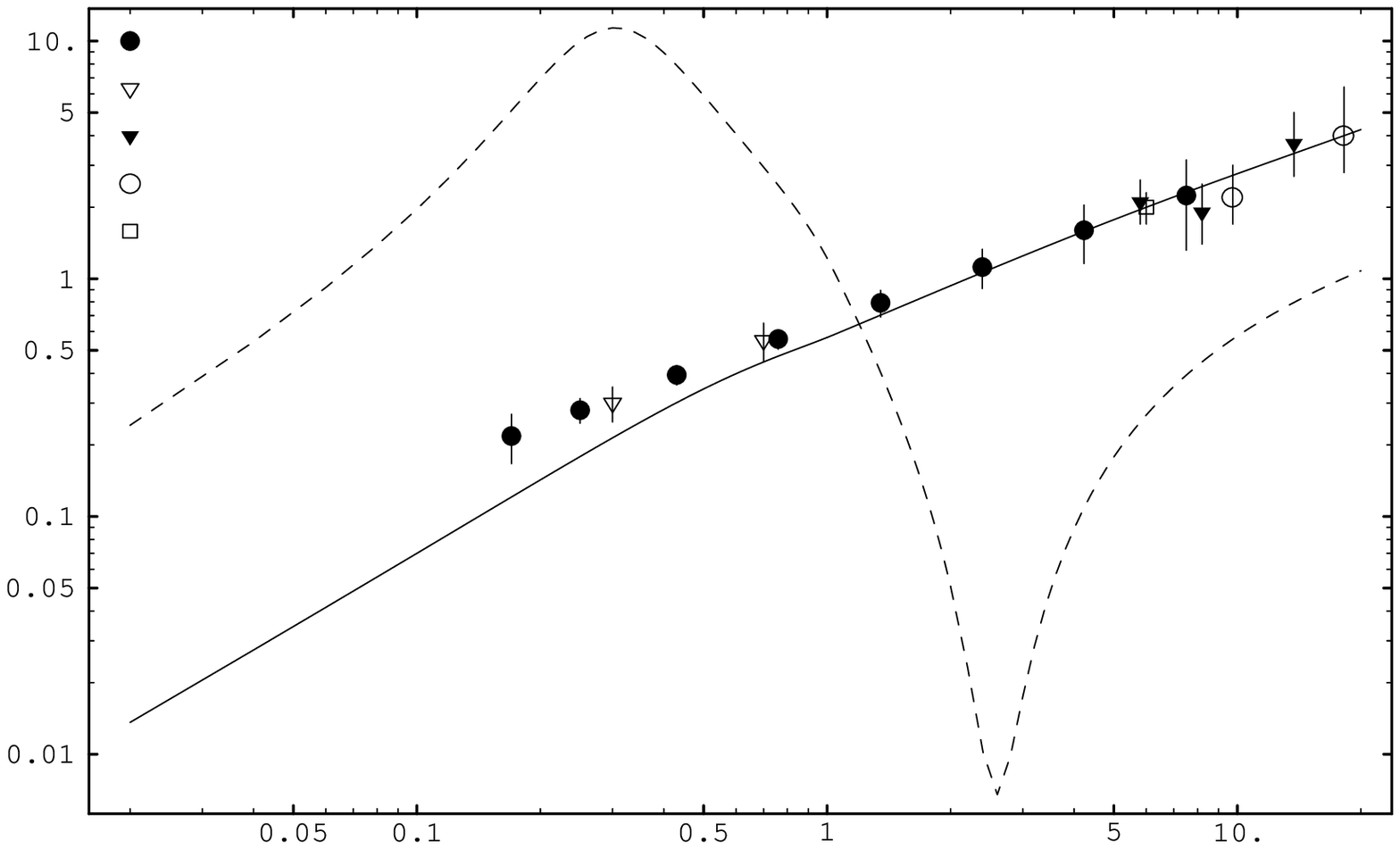}
\begin{picture}(0,0)
\put(-4,0){\makebox(0,0){$Q^2[\!\!\GeV^2]$}}
\put(-104,95){\makebox(0,0){$R_{LT}(Q^2)\!=\!\si^L(Q^2)\!/\!\si^T(Q^2)$}}
\put(-119,17){\makebox(0,0){$\rh$}}
\put(-117,47){\makebox(0,0){$2S$}}
\put(-118,85.75){\makebox(0,0){\scriptsize E665}}
\put(-95.5,81){\makebox(0,0){\scriptsize ZEUS BPC 1995 prel.}}
\put(-103,76.25){\makebox(0,0){\scriptsize ZEUS 1994 prel.}}
\put(-107,71.5){\makebox(0,0){\scriptsize H1 1994 prel.}}
\put(-117.5,66.75){\makebox(0,0){\scriptsize NMC}}
\end{picture}
$$
\vspace*{-6ex}
\caption[]{\small Ratio of longitudinal to transverse integrated cross sections as function of $Q^2$ both for the $\rh$-meson and the $2S$-state. There is only data for $\rh$-production~\cite{E665_1}. } \label{Fig:R_LT}
\end{figure}
In Fig.~\ref{Fig:R_LT} we display the ratio $R_{LT}(Q^2)$ of longitudinal to transverse coss sections and find good agreement with experimental data for the $\rh$-state. For the $2S$-state we again predict a marked structure which is very sensitive to the node positions in the wave functions.

Further results refering to cross sections differential in~$-t$ and the ratio of $2\pi^+2\pi^-$-production via $\rh'$ and $\rh''$ to $\pi^+\pi^-$-pro\-duction via~$\rh$ are given in Ref.~\cite{KDP}.

\begin{center} {\large\it Ackknowledgements} \end{center}
The author thanks H.G.~Dosch and H.J.~Pirner for collaboration in the underlying work.

\renewcommand{\refname}{{\begin{center}\large\it References\end{center}}}
\vspace*{0ex}   

\end{document}